\documentclass[conference,letterpaper]{IEEEtran}

\usepackage[pdftex]{graphicx}
\graphicspath{{../pdf/}{../jpeg/}}
\DeclareGraphicsExtensions{.pdf,.jpeg,.png}
\usepackage[utf8]{inputenc} 
\usepackage[T1]{fontenc}
\usepackage{url}
\usepackage{ifthen}
\usepackage{cite}
\usepackage{amssymb}
\usepackage{subcaption}
\usepackage[cmex10]{amsmath} 

\begin{document}
\title{The Filtered Gaussian Primitive Diamond Channel}


\author{%
  \IEEEauthorblockN{Asif~Katz}
  \IEEEauthorblockA{Technion - Israel Institute\\ of            Technology\\
          asifk@campus.technion.ac.il}
  \and
  \IEEEauthorblockN{Michael~Peleg}
  \IEEEauthorblockA{Technion - Israel Institute\\ of                       Technology\\
                    and Rafael ltd\\
                    peleg.michael@gmail.com}
  \and
  \IEEEauthorblockN{Shlomo~Shamai~(Shitz)}
  \IEEEauthorblockA{Technion - Israel Institute\\ of                       Technology\\
                    sshlomo@ee.technion.ac.il}
}


\maketitle

\begin{abstract}
We investigate the special case of diamond relay comprising a Gaussian channel with identical frequency response from the user to the relays and fronthaul links with limited rate from the relays to the destination. We use the oblivious compress and forward (CF) with distributed compression and decode and forward (DF) where each relay decodes the whole message and sends half of its bits to the destination. We derive achievable rate by using time-sharing between DF and CF. It is proved that optimal CF-DF time sharing is advantageous over superposition of CF and DF.
The optimal time sharing proportion between DF and CF and power and rate allocations are different at each frequency and are fully determined.
\end{abstract}
\section{Introduction}
\IEEEPARstart{R}{elaying} is a classical technique in communications systems, which is of theoretical and practical importance.
It is the central element in the wireless cell-free technology, where the complete decoding is performed only at the final destination. Indeed the distributed non-cooperative relaying is the basic elements in what is known as the Cloud Radio Access Network (CRAN) \cite{fronthaulcompressionCRAN}, \cite{CapacityCRAN}, where there are several relays, each possesses an error free, fronthaul link to a cloud computing central processor. Another example, of central practical implications of such a scheme is based on remote radio heads systems connected to base stations with common public radio interface \cite{cpriCRAN}. Our study is directly associated with these models, where the focus is on point-to-point communications, implying the primitive diamond channel. Further, we extend the view of classical oblivious processing, based on distributed CF as examined in \cite{our_prev} and allow the relays to combine CF and DF in an optimal way. It is shown that the optimal combination is time-sharing between CF and DF, that is in our case, dividing time and frequency assignments between CF and DF. In the optimal solution, DF must clearly comply exactly with classical water-filling, and CF must comply with the rules presented in \cite{our_prev}. The combination of CF and DF using the randomized time-sharing strategy was shown in \cite{gamal21} to improve the performance for the single relay channel.
System rate optimization using Lagrange multipliers and Karush–Kuhn–Tucker (KKT) conditions for various problems was also studied in \cite{xing2020new}. However, it does not include the diamond relay system, which requires also relay rates constraints in addition to the applied power constraints.
Preliminary results of this work will be preseted in \cite{Katz2021}.
In this section we next describe the system model and the information bottleneck method. In Section \ref{section_preliminaries} we review the information bottleneck problem and known previous results of upper bounds and achievable rates for the discrete time model. We then describe the time sharing approach between CF and DF and the superposition coding approach. In Section \ref{superposition_section} we analyze the optimal solution of superposition coding and prove that in our system model time sharing has equal or better better performance than superposition coding. In Section \ref{section_flat_frequency} we investigate the flat frequency response optimal solution using CF and DF time sharing and compare the optimal rate of our scheme to the upper bounds and achievable rates described in Section \ref{section_preliminaries}. In Section \ref{section_frequency_selective} we extend the time sharing optimization to the frequency-selective case and compare its performance to previous results. In Section \ref{optimal_solution_section} we further analyze and prove properties of the behavior of the optimal solution for the frequency selective case. Section \ref{section_conclusions} concludes the paper.
\subsection{System model}
\begin{figure}[ht]
\begin{subfigure}{0.5\textwidth}
    \centering
    \includegraphics[width=\textwidth]{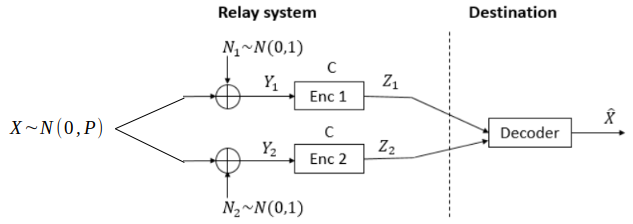}
    \caption{Gaussian diamond relay channel scheme for the discrete time frequency-flat case}
    \label{system_discrete}
\end{subfigure}
\begin{subfigure}{0.5\textwidth}
    \centering
    \includegraphics[width=\textwidth]{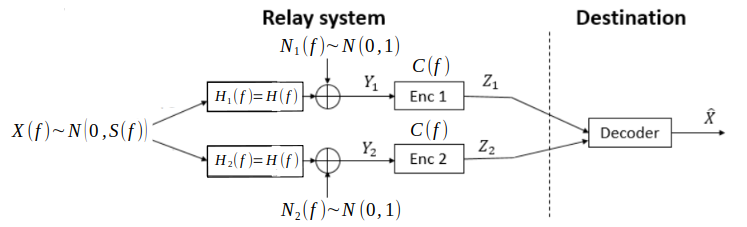}
    \caption{Gaussian diamond relay channel scheme for the frequency selective case}
    \label{system_freq}
\end{subfigure}
\caption{Gaussian diamond relay channel scheme}
\label{system_schemes}
\end{figure}
The system models for the discrete time and the frequency selective cases are shown in Fig. \ref{system_discrete} and Fig. \ref{system_freq} respectively. A real Gaussian signal X is transmitted over two additive white Gaussian noise (AWGN) relay channels. Each relay channel has signal to noise ratio (SNR) equal to $P$ in the frequency-flat case. For the frequency selective case the noise one-sided power spectral density is unity and the channel response affects the SNR as the signal $X(f)$ is multiplied by the frequency-dependent filter response. Each relay has a rate limited encoder connected to the destination decoder via a fronthaul link Z. In this paper we limit the model to the case where $H_1(f) = H_2(f) = H(f)$. Each relay channel has limited bit rate C[bits/ sec] fronthaul channel from the encoder to the decoder at the destination. The relay encoders, as pointed out, do time sharing between CF and DF, and they do not communicate with each other. We aim to maximize the mutual information between the source X and the destination $\hat{X}$ subject to the source power constraint and a fronthaul rate constrained link between the relays and the destination. 
\section{Preliminaries}
\label{section_preliminaries}
In this section we summarize previous results of discrete time frequency-flat Gaussian diamond relay channel, as shown in Fig. \ref{system_discrete}. The transmitter uses classic codes, for real Gaussian-distributed X, and the channel to each relay is AWGN.
\subsection{Information bottleneck}
The Information Bottleneck (IB) method \cite{tishby99information} can be used in order to find an optimal mapping according to a balance between maximizing mutual information of source and destination, while being constrained by the relay to destination rate. For the oblivious system, that is, a system in which the relay is oblivious of the error correction codes used, with a single relay, this method yields the optimal solution. An extension for this method, named distributed bottleneck, which is related to distributed compression \cite{Sanderovich2008}, \cite{zaidi2019} was shown to be optimal for the oblivious case in \cite{CapacityCRAN}. This result was also obtained for the classical CEO problem with logarithmic distortion in \cite{ugur2020vector}, \cite{aguerri2017distributed}. To achieve an information throughput approaching the mutual information, a standard error correcting code, external to the system analyzed here, is used. For the non-oblivious case, DF can be used by having the relays decoding
the messages, functioning as receivers in a broadcast channel. We showed the optimal rate of distributed CF over frequency-selective
channels in \cite{our_prev} and the methodology is used in this paper.
\subsection{Compress and forward}
In this transmission method, in each transmission block the relay quantizes its received message, encodes and transmits it to the destination in the following transmission block. In this method the relays can be oblivious about the encoding scheme, which gives the system various advantages which are discussed in \cite{our_prev}.
The system rate $R_{CF}$ in $[bits/channel \ use]$ when using CF with joint decompression and decoding was shown in \cite{Sanderovich2008} and is based on distributed bottleneck as in \cite{CapacityCRAN}.
Noisy network coding described in \cite{NoisyNetworkCoding}, was shown in \cite{CapacityCRAN} to be equivalent to the oblivious CF processing, along with the above method. We use Eq. (\ref{cf_rate}) that was derived in \cite{Sanderovich2008} with our calculation. We define here the SNR of each AWGN relay channel as $P_{CF}$ and the fronthaul channel rate limit in bits per channel use as $C_{CF}$.
\begin{flalign}
\label{cf_rate}
R_{CF} =& \frac{1}{2}log_2\left[1+2P_{CF}\cdot2^{-4C_{CF}}\cdot{\left(2^{4C_{CF}}+P_{CF}\right.}\right.\\\nonumber
&\left.{\left.-\sqrt{P_{CF}^2+(1+2P_{CF})\cdot2^{4C_{CF}}}\right)}\right]
\end{flalign}
Our system model was investigated in \cite{ganguly2019capacity}, which derived a rate equation that coincides with the rate in Eq. (\ref{cf_rate}). 
\begin{flalign}
\label{cf_rate_sigma}
&R_{CF} = max_{\sigma>0} R_{CF}(\sigma)&&\\\nonumber
\\\nonumber
&R_{CF}(\sigma) = min\left( \frac{1}{2}log_2(1+\frac{2P_{CF}}{1+\sigma^2}),\right.\\\nonumber&\left.\frac{1}{2}log_2(1+\frac{P_{CF}}{1+\sigma^2})+C_{CF}-\frac{1}{2}log_2(1+\frac{1}{\sigma^2}),\right.
\\\nonumber&
\left.2C_{CF}-log_2(1+\frac{1}{\sigma^2}) \right)
\end{flalign}
Here the minimum is over all possible oblivious relays cut-sets, similar to the general relays cut-set bound. $\sigma^2$ is the variance of a zero mean Gaussian noise added to the received signal at each relay caused by the compression. Lower $\sigma^2$ values refer to less logarithmic distortion, which is preferred when the relay to destination rate is high, while higher $\sigma^2$ values are where enhanced distortion results, this is preferred where the relay to destination rate is low. For optimal frequency allocation solution we use the results from \cite{our_prev} as discussed in Appendix \ref{DF_CF_appendix}.
\subsection{Decode and forward}
In this transmission method, each relay decodes its message and sends it to the destination through a fronthaul noiseless link.
The DF rate for our system is the known Gaussian broadcast channel capacity with $SNR_1=SNR_2=P_{DF}$
\begin{equation}
\label{df_rate}
R_{DF} =  \frac{1}{2}log_2(1+P_{DF}) [bits/channel \ use]
\end{equation}
and the required fronthaul channel rate is \\ $C_{DF} \geq \frac{R_{DF}}{2} = \frac{1}{4}log_2(1+P_{DF}) \ [bits/channel \ use]$ because each relay is required to send half of the message to the destination.
Optimal frequency allocation solution is derived in Appendix \ref{DF_CF_appendix}.
\subsection{Time sharing}
\begin{figure}[htbp]
\centering
\includegraphics[width=0.5\textwidth]{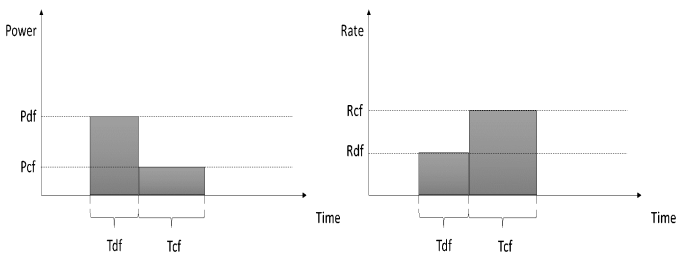}
\caption{Time sharing between DF and CF}
\label{time_sharing_example}
\end{figure}

In our scheme we use time sharing with both CF and DF. In the first phase, both relays transmit DF over time $T_{DF}$ with power $P_{DF}$ and fronthaul rate $C_{DF}$. In the second phase both relays transmit CF over time $T_{CF}$ with power $P_{CF}$ and fronthaul rate $C_{CF}$.
The time sharing is described in Fig. \ref{time_sharing_example}, where $P_{DF}>P_{CF}$ and $R_{DF}<R_{CF}$. At the first time portion DF is used at the relays, and at the second time portion CF is used. The time allocation is done according to the total average power and rate constraints. Further analysis is done in Sections \ref{section_flat_frequency} and \ref{section_frequency_selective}.
\subsection{Superposition coding}
In the superposition approach, the transmitter will transmit DF letter along with CF letter. Each relay will receive the DF and CF letters with added noise. The DF letter will be decoded at the relay and then transmitted to the destination. This letter will also be subtracted from the relay input, so the remaining CF letter with the channel noise will next be transmitted to the destination, which will then decode the CF letter from the information received from both relays. As can be seen from this scheme, CF operation will remain the same, because DF wont affect its compress and forward operation as it is subtracted before the compression. However, DF operation is affected from the CF letter that acts as an additional noise to the relay decoder.
The system rate $R [bits/channel \ use]$ for the superposition coding (SPC) scheme is
\begin{flalign}
\label{spc_rate}
R_{SPC}=R_{DF}\left(\frac{P_{DF}}{1+P_{CF}}\right)+R_{CF}(P_{CF},C_{CF})
\end{flalign}
where DF and CF rate functions here are the same as written in Eq. (\ref{df_rate}) and (\ref{cf_rate}) respectively.
\subsection{Upper bounds}
For the general diamond relay channel, the cut-set upper bound for the system rate $R[bits/channel \ use]$ is a classical result shown in \cite{elgamal_kim_2011}. The rate region is
\begin{flalign}
\label{cut_set_equations}
&R \leq I(X;Y_1,Y_2) \\\nonumber
&R \leq I(X;Y_1)+C_2 \\\nonumber
&R \leq I(X;Y_2)+C_1 \\\nonumber
&R \leq C_1+C_2
\end{flalign}
In our Gaussian model with $SNR_1=SNR_2=P$ and $C_1=C_2=C [bits/channel \ use]$ the bound becomes
\begin{flalign} 
\label{cutset_rate}
R_{cutset} = min\left(\frac{1}{2}log_2(1+2P),\frac{1}{2}log_2(1+P)+C,2C\right)
\end{flalign}
In this paper we compare our scheme results to the cut-set upper bound at Eq. (\ref{cutset_rate}), and to a tighter upper bound shown in \cite{newupperbound}, that accounts for the tension between information measures in relevant Markov chains.
\section{Superposition coding analysis}
\label{superposition_section}
In this section we will investigate the superposition coding approach for our system and compare its performance to the time-sharing approach. 
We first write the power and rate constraints using the discrete time equations
\begin{flalign}
\label{spc_constraints}
P_{DF}+P_{CF}=P \\\nonumber  
C_{DF}+C_{CF}=C
\end{flalign}
Using (\ref{spc_constraints}) we get the power constraint $P_{CF} = P - P_{DF}$ and rate constraint $C_{CF} = C - C_{DF}$.
Here as before, we also assume that when DF is used, it will be at its optimal point where
\begin{flalign}
\label{spc_cdf}
C_{DF} = \frac{1}{2}R_{DF}\left(\frac{P_{DF}}{1+P_{CF}}\right)=\frac{1}{2}R_{DF}\left(\frac{P_{DF}}{1+P-P_{DF}}\right)
\end{flalign}
\\
Thus (\ref{spc_rate}) can be written as a function of $P_{DF}$, so given system constraints P and C we get the following optimization problem 
\begin{flalign}
\label{optimization_problem}
&max_{P_{DF}}\ R_{SPC}(P_{DF})\\\nonumber
&0 \leq P_{DF} \leq P \\\nonumber
&0 \leq C_{DF} \leq C
\end{flalign}
\\
The following figure shows the optimal solution as a function of the relay rate $C$ for $P=3$
\begin{figure}[htbp]
\centering
\includegraphics[width=0.5\textwidth]{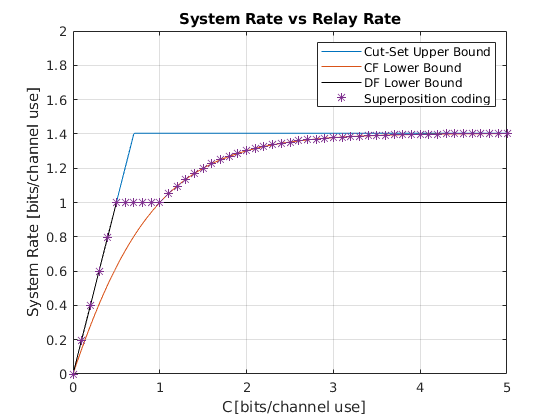}
\caption{Gaussian diamond relay system rates with superposition}
\label{system_bounds_superposition}
\end{figure}
\\
As can be seen in Fig. \ref{system_bounds_superposition}, the optimal solution for superposition coding will assign resources to either DF or CF - the one with the higher rate. The following theorem covers the general case.\\
\textbf{Theorem 1:} Given communication over the diamond relay channel using combination of DF and CF by superposition as presented above and with any parameters, there always exists a CF only or DF only scheme which performs at
least as well.\\
\textbf{Proof:} We first write the rate function as described in Eq. (\ref{spc_rate}), (\ref{spc_constraints}), (\ref{spc_cdf})
\begin{flalign}
\label{spc_rate_pdf}
R_{SPC}=&R_{DF}\left(\frac{P_{DF}}{1+P-P_{DF}}\right) \\ \nonumber
&+R_{CF}\left(P-P_{DF},C-\frac{1}{2}R_{DF}\left(\frac{P_{DF}}{1+P-P_{DF}}\right)\right)
\end{flalign}
In order to prove the theorem, we will investigate the first derivative of the rate function as a function of $P_{DF}$. Next we write $R_{DF}=R_{DF}(SNR)=R_{DF}\left(\frac{P_{DF}}{1+P-P_{DF}}\right)$.
Using the chain rule we can obtain the rate function first derivative
\begin{flalign}
\frac{dR_{SPC}}{dP_{DF}} = \frac{dR_{DF}}{dP_{DF}}+\frac{dR_{CF}}{dP_{CF}}\cdot \frac{dP_{CF}}{dP_{DF}} +\frac{dR_{CF}}{dC_{CF}}\cdot \frac{dC_{CF}}{dP_{DF}} \nonumber
\end{flalign}
Using that $P_{CF}=P-P_{DF}$ and $C_{CF}=C-C_{DF}=C-\frac{R_{DF}}{2}$ we get
\begin{flalign}
\label{derivative_rspc}
\frac{dR_{SPC}}{dP_{DF}} =& \frac{dR_{DF}}{dP_{DF}}-\frac{dR_{CF}}{dP_{CF}}\bigg\rvert_{P_{CF}=P-P_{DF}} \\\nonumber &-\frac{1}{2}\cdot\frac{dR_{DF}}{dP_{DF}}\cdot\frac{dR_{CF}}{dC_{CF}}\bigg\rvert_{C_{CF}=C-\frac{R_{DF}}{2}}\nonumber
\end{flalign}
For this continuous function, if an inner point where $0<P_{DF}<P$ would be optimal, its first derivative would be zero and the following equation is fulfilled 
\begin{flalign}
\label{derivative_equation}
\frac{dR_{DF}}{dP_{DF}}\cdot\left[1-\frac{1}{2}\cdot\frac{dR_{CF}}{dC_{CF}}\right]=\frac{dR_{CF}}{dP_{CF}}
\end{flalign}
\\
For a given power $P$ and any $0 \leq P_{DF} \leq P$ the only solution of $C$ for Eq. (\ref{derivative_equation}) yields
\begin{flalign}
\label{rate_th}
C=C_{SPC-TH}=\frac{1}{4}log_2(1+P)+\frac{1}{2}
\end{flalign}
For a given power constraint $P$, the derivative (\ref{derivative_rspc}) is a continuous function of $C$ and $P_{DF}$ so in each region of $C>C_{SPC-TH}$ and $C<C_{SPC-TH}$ its value is either negative or positive for every $P_{DF}$ value and does not change sign inside the region. Thus, the optimal solution would be either $P_{DF}=0$ for negative derivative or $P_{DF}=P$ for positive derivative. Substituting certain points shows that for $C>C_{SPC-TH}$ the derivative is negative and for $C<C_{SPC-TH}$ it is positive. This means that for $C>C_{SPC-TH}$ CF will be chosen as optimal solution and for $C<C_{SPC-TH}$ DF will be chosen as optimal solution. The rate $C_{SPC-TH}$ is also the point where for equal power constraint, only DF and only CF schemes achieve the same system rate. Therefore the optimal solution for superposition coding will be either DF or CF and the transition is where their rates are equal for same power allocation, making the total rate to be the maximum between them. This behavior is expected - DF is preferred at low rate constraint and CF is preferred at high rate constraint. For $C=1$ the region dividing line is at $P=3$, which is expected as the point of switch between DF and CF for those constraints for the superposition scheme, as shown in Fig. \ref{system_bounds_superposition}. However, we observe that the equation holds until a minimum relay rate, thus supports relay rate values of $C \geq \frac{1}{2}$.
The following Fig. \ref{first_derivative_regions} shows the first derivative regions for several power constraints, where below the region lines the first derivative is positive and above them it is negative. The dashed line is $C=C_{SPC-TH}$ for $P=P_{DF}$. For a given power constraint, at higher relay rates CF will be chosen and for lower rates DF will be chosen.
\begin{figure}[!htbp]
\centering
\includegraphics[width=0.5\textwidth]{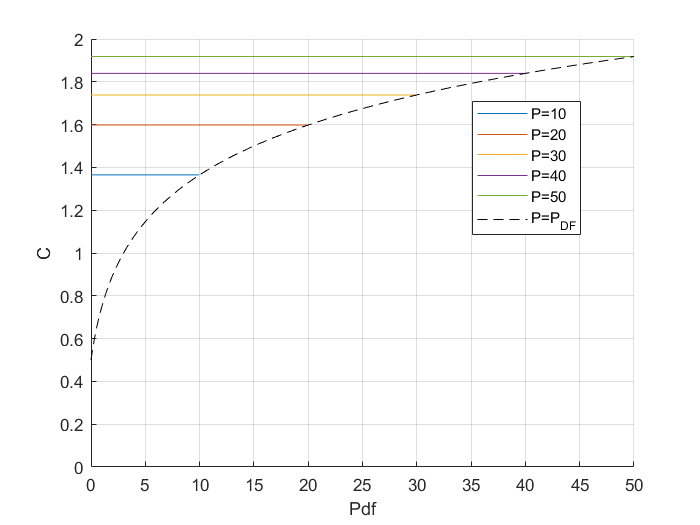}
\caption{First derivative regions lines for different P values}
\label{first_derivative_regions}
\end{figure}
The CF only and DF only schemes could be obtained from the time-sharing approach, which is a more general case of them. Therefore, time-sharing has equal or better performance than CF only and DF only, so by Theorem 1 it has also equal or better performance than superposition coding.
\section{Flat frequency response analysis}
\label{section_flat_frequency}
We now investigate the diamond relay system shown in
Fig. \ref{system_freq} for the flat frequency response case. The transmitter uses classic codes, for real Gaussian-distributed X. In this case the channel response is uniform over the bandwidth from zero to $W=\frac{1}{2} [Hz]$ and is set to be $H_1(f)=H_2(f)=1$, so it does not affect X, while the channel to each relay is AWGN. When using CF the compression in each relay is done by using remote source-coding with distributed compression. When using DF each relay decodes the message and transmit half of the bits to the destination. For our frequency-flat case the optimization of allocating DF and CF in separate frequency bands can also be calculated using the discrete time system time-sharing optimization. We next use this approach so the frequency-selective system shown in Fig. \ref{system_freq} is now simplified as shown in Fig. \ref{system_discrete}. 
\subsection{System Rates}
\begin{figure}[htbp]
\centering
\includegraphics[width=0.5\textwidth]{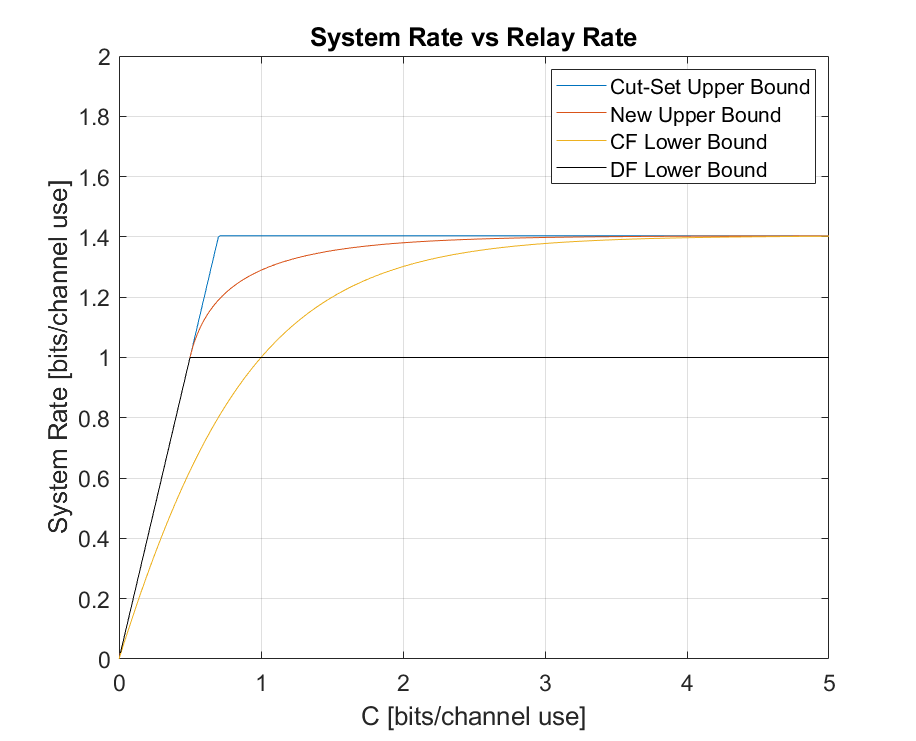}
\caption{Gaussian diamond relay system rates without time-sharing, $P=3$}
\label{system_bounds}
\end{figure}

For the flat frequency response case with $W=\frac{1}{2} [Hz]$ and $H_1(f)=H_2(f)=1$ we can readily infer from Section \ref{water-pouring subsection} that the rates equations equal the discrete time equations, so the cut-set upper bound of the diamond relay system is as in Eq. (\ref{cutset_rate}), the CF rate is as in Eq. (\ref{cf_rate}) and DF rate is as in Eq. (\ref{df_rate}).
Fig. \ref{system_bounds} shows the above rates as a function of the relay to destination rate $C$ and $P=3$.
It can be seen that for low relay rate the DF system rate is higher while for high relay rates the CF system rate is higher.
Here the CF rate of Eq. (\ref{cf_rate_sigma}) coincides with the rate in Eq. (\ref{cf_rate}). We can immediately infer that using a simple switch between CF and DF and choosing the better one for a certain relay rate would have a better performance than using only one of them.
\subsection{Time sharing}
\begin{figure}[htbp]
\centering
\includegraphics[width=0.5\textwidth]{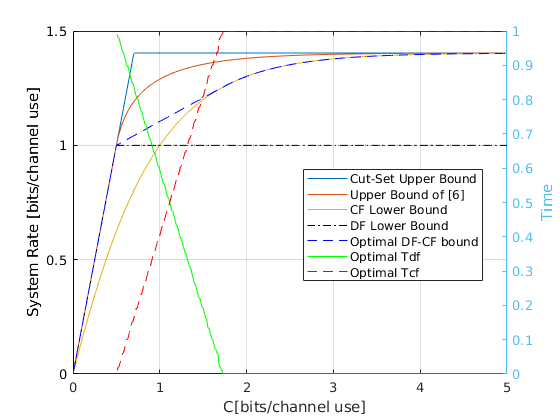}
\caption{Gaussian diamond relay system upper bounds and CF, DF rates with and without time-sharing for various relay rates, $P=3$}
\label{system_rate_timesharing}
\end{figure}

\begin{figure}[htbp]
\centering
\includegraphics[width=0.5\textwidth]{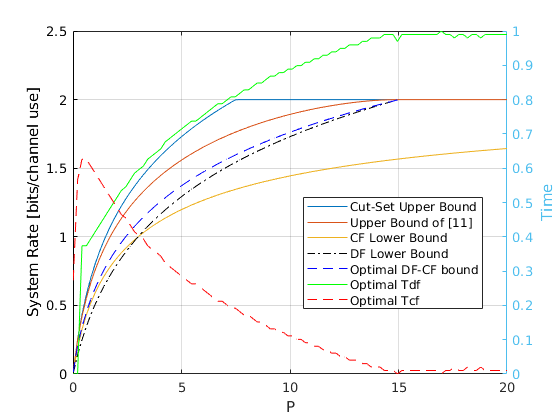}
\caption{Gaussian diamond relay system upper bounds and CF, DF rates with and without time-sharing for various powers, $C=1$}
\label{system_rate_timesharing_power}
\end{figure}

Now we will investigate the optimal solution of the time sharing scheme described in Section \ref{section_preliminaries}.
As seen in Fig. \ref{system_rate_timesharing}, an optimal solution of DF with a given $P_{DF}$ will allocate the minimal required relay rate $C_{DF}$ that achieves the DF rate at its time slot, because increasing it would not increase the system rate. As shown in section \ref{section_preliminaries}, the minimal rate can be calculated directly from DF allocated power, therefore $C_{DF}$ would be a function of $P_{DF}$ and not a variable of the optimization problem.
Now we will write the time sharing optimization problem for the flat frequency response case
\begin{flalign}
\label{flat_frequency_optimization}
&\underset{P_{DF},P_{CF},C_{CF},T_{CF},T_{DF}}{max}\ T_{DF} \cdot R_{DF} + T_{CF} \cdot R_{CF}\\
\nonumber&s.t.\\
\nonumber&0 \leq T_{DF} \leq 1\\
\nonumber&0 \leq T_{CF} \leq 1\\
\nonumber&0 \leq T_{DF}+T_{CF} \leq 1\\
\nonumber&0 \leq T_{DF} \cdot P_{DF} + T_{CF} \cdot P_{CF} \leq P\\
\nonumber&0 \leq T_{DF} \cdot C_{DF} + T_{CF} \cdot C_{CF} \leq C\\
\nonumber&C_{DF} = \frac{R_{DF}}{2} = \frac{1}{4}log_2(1+P_{DF})
\end{flalign}
The rates in Eq. (\ref{flat_frequency_optimization}) are those written in Eq. (\ref{cf_rate}) and Eq. (\ref{df_rate}). The optimal solution for both DF and CF allocation is shown in Appendix \ref{DF_CF_appendix} for the frequency-selective case. The flat frequency response model optimization can be solved by simpler two variable grid search, which we used in order to verify our proposed method results. 
The optimal solution as a function of the relay to destination rate $C$ with power constraint $P=3$ is shown in Fig. \ref{system_rate_timesharing}. Solution is obtained using the Lagrange multipliers method described in \cite{our_prev} and also in Section \ref{section_frequency_selective} for the frequency-selective case. We use MATLAB Symbolic Toolbox in order to find the analytic solutions expressions and the Lagrange multipliers region, then we use Optimization Toolbox to find the optimal solution. It can be seen that at low relay rates the DF part is dominant. As the relay rate increases, the system rate increases and Tdf decreases, which indicate that the CF stage becomes relevant as it can use high relay rate and increase the total channel rate. This is until the optimal rate coincides with the CF rate. From this behavior we can infer that CF consumes more link rate resources than DF and therefore is used only when there is enough excess rate than using only DF.
We applied the same method in order to find optimal solution as a function of the power constraint $P$ with relay to destination rate $C=1$, which is shown in Fig. \ref{system_rate_timesharing_power}. At very low power only CF is allocated and only for part of the total time. It is expected that CF will be preferred in this region as it is known, for example from \cite{Cooperative_Communications} that for the single relay channel CF is preferred over DF in the case where the relays are far from the source, which can be thought as low power case in our system. As the power increases, DF is being allocated so we have time-sharing between CF and DF. $T_{CF}$ increases until it achieves maximum value, then it decreases until reaching $T_{DF}$ and after that it decreases to zero. From this behavior we can infer that DF consumes more power resources than CF and therefore is used only when there is enough power. From this figure we can infer that using time-sharing will improve the system rate and the largest improvement is in the case where the power value is at the medium values range, this is where the relays are at medium range from the source. When the relays are close to the source the time-sharing will prefer DF for best performance, and when they are very far from the source it will prefer CF.  

\section{Frequency-selective case analysis}
\label{section_frequency_selective}
We now investigate the diamond relay system shown in
Fig. \ref{system_freq} for the frequency selective case. In this case the channel response is set to be $H_1(f)=H_2(f)=H(f)$. In addition to the channel response there is an AWGN at each relay. Each relay uses time sharing between CF and DF as described in Section \ref{section_flat_frequency}.
\subsection{Generalized Water-pouring}
\label{water-pouring subsection}
In this paper we use the same generalized water-pouring approach that was explained and used in \cite{our_prev}. We derive DF solution in Appendix \ref{DF_CF_appendix} and show that the additional time-sharing variables do not affect the optimal solution. This allows us to use previous results for CF, shown in \cite{our_prev}.
To apply the results from the real-valued frequency-flat channel to the frequency-selective channel we decompose the channel, as in standard water pouring, into infinitesimally small frequency bands of $df[Hz]$ where the total bandwidth is $W[Hz]$, e.g. \cite[Chapter 7]{lapidoth_2017}, \cite{SHANNON1949} and \cite{Tsybakov1970}. In each $df$ band we have $2 \cdot df$ real channel uses per second. The signal power in each $df$ band is $S(f) \cdot df$, which leads to $SNR=S(f)$ for each channel use and in each band. The signal power equation for the frequency-selective case is
\begin{flalign}
P = \int_0^W S(f) \cdot df
\end{flalign}
where $P$ is in $[Watt]$ and $S(f)$ is in $[Watt/Hz]$.
The system rate $R [bits/sec]$ is
\begin{flalign}
R = \int_0^W 2R(f) \cdot df
\end{flalign}
where $R(f)$ denotes the number of bits transferred per one real-channel use. The rate per bandwidth in $[bits/(sec \cdot Hz)]$ is then $2\cdot R(f)$.
Similarly, $C(f)$ denotes twice the number of bits used by each relay per one real-channel use yielding
\begin{flalign}
C = \int_0^W \frac{1}{2}C(f) \cdot 2df = \int_0^W C(f) \cdot df
\end{flalign}
where $C$ is in $[bits/sec]$ and $C(f)$ is in $[bits/(sec \cdot Hz)]$.

\subsection{System rates}
We can now generalize the optimization problem of the frequency-selective case. Based on the definition above, the rates are now derived for the frequency-selective case as $[bits/(sec \cdot Hz)]$, so integrating over the frequency band will result in $[bits/sec]$.
We first derive DF rate in $[bits/(sec \cdot Hz)]$ using Eq. (\ref{df_rate}). The power at each frequency $S_{DF}(f)$ is affected by the channel frequency response $H(f)$ so the rate becomes
\begin{equation}
\label{df_rate_frequency}
R_{DF}(f) =  \frac{1}{2}log_2\left(1+S_{DF}(f)\cdot|H(f)|^2\right)
\end{equation}
and by assigning fronthaul rate per bandwidth of $\frac{1}{2}C_{DF}(f)$ to each real-channel use we get the constraint $C_{DF}(f) \geq R_{DF}(f) = \frac{1}{2}log_2(1+S_{DF}(f)\cdot|H(f)|^2) \ [bits/(sec \cdot Hz)]$. As discussed above the optimal DF fronthaul rate is the minimal value which is a function of $S_{DF}(f)$ and $H(f)$.\\
Similarly we can derive the CF rate in $[bits/(sec \cdot Hz)]$ using Eq. (\ref{cf_rate})
\begin{flalign} 
\label{cf_rate_frequency}
&R_{CF}(f) =  \frac{1}{2}log_2\left[1+2A(f)\cdot2^{-2C_{CF}(f)}\cdot{\left(2^{2C_{CF}(f)}+A(f)\right.}\right.\\\nonumber&\left.{\left.-\sqrt{A(f)^2+(1+2A(f))\cdot2^{2C_{CF}(f)}}\right)}\right]
\\ \nonumber &A(f) \triangleq S_{CF}(f)\cdot|H(f)|^2
\end{flalign}
\subsection{Time sharing}
Now we can write the frequency-selective optimization problem of the total system rate using time sharing between CF and DF
\begin{flalign}
\label{multi_freq_optimization}
&max_{\substack{S_{DF}(f)\\S_{CF}(f)\\C_{CF}(f)\\T_{DF}(f)\\T_{CF}(f)}} \int_0^W [T_{DF}(f) \cdot R_{DF}(f) + T_{CF}(f) \cdot R_{CF}(f)] \cdot 2df \\
\nonumber&s.t.\\
\nonumber&0 \leq T_{DF}(f) \leq 1\\
\nonumber&0 \leq T_{CF}(f) \leq 1\\
\nonumber&0 \leq T_{DF}(f)+T_{CF}(f) \leq 1\\
\nonumber&0 \leq \int_0^W [T_{DF}(f) \cdot S_{DF}(f) + T_{CF}(f) \cdot S_{CF}(f)] \cdot df \leq P\\
\nonumber&0 \leq \int_0^W [T_{DF}(f) \cdot C_{DF}(f) + T_{CF}(f) \cdot C_{CF}(f)] \cdot df \leq C\\
\nonumber&C_{DF}(f) = R_{DF}(f) = \frac{1}{2}log_2(1+S_{DF}(f)\cdot|H(f)|^2)
\end{flalign}
In Appendix \ref{DF_CF_appendix} we show that $T_{DF}(f)$ and $T_{CF}(f)$ values does not affect the gradient. Therefore, we first calculate optimal solution of $S_{DF}(f),S_{CF}(f),C_{CF}(f)$ using the gradient and then calculate optimal $T_{DF}(f)$ and $T_{CF}(f)$ values according to the constraints. We choose the same CF solution as in \cite{our_prev} so $R_{CF}(f)$ is concave with respect to $S_{CF}(f), C_{CF}(f)$. $R_{DF}(f)$ is concave with respect to $S_{DF}(f)$ as the known logarithm function. Because $R_{DF}(f)$ does not depend on $S_{CF}(f), C_{CF}(f)$ and $R_{CF}(f)$ does not depend on $S_{DF}(f)$, they both concave with respect to $S_{DF}(f), S_{CF}(f), C_{CF}(f)$. Therefore linear combination of $R_{CF}(f)$ and $R_{DF}(f)$ is also concave with respect to $S_{DF}(f), S_{CF}(f), C_{CF}(f)$. Therefore, our total system rate is concave function of $S_{DF}(f),S_{CF}(f),C_{CF}(f)$ and the optimization problem can be solved by the Lagrange multipliers method, which is similar to \cite{our_prev}.
The Lagrangian function is
\begin{flalign}
\label{lagrangian}
&L(f,S_{DF}(f),S_{CF}(f),C_{CF}(f),\lambda_C,\lambda_S) = \\
\nonumber&\int_0^W [T_{DF}(f) \cdot R_{DF}(f) + T_{CF}(f) \cdot R_{CF}(f)] \cdot 2df\\
\nonumber&-\lambda_S\left[\int_0^W[T_{DF}(f) \cdot S_{DF}(f) + T_{CF}(f) \cdot S_{CF}(f)] \cdot df - P\right]\\
\nonumber&-\lambda_C\left[\int_0^W [T_{DF}(f) \cdot C_{DF}(f) + T_{CF}(f) \cdot C_{CF}(f)] \cdot df - C\right]
\end{flalign}
This is because $C_{DF}(f)$ is a function of $S_{DF}(f)$ so the Lagrangian function variables are $S_{DF}(f),S_{CF}(f),C_{CF}(f),\lambda_C,\lambda_S$. According to the KKT theorem we would find a saddle point of the Lagrangian function which is also an optimal point of our optimization problem. The gradient of the Lagrangian function for $S_{DF}(f),S_{CF}(f),C_{CF}(f)$ is
\begin{equation}
    \label{gradient_lagrangian}
    \nabla{L}=\left(\frac{dL}{dS_{DF}(f)},\frac{dL}{dS_{CF}(f)},\frac{dL}{dC_{CF}(f)}\right)(f)
\end{equation}
We know that frequency selective optimal solutions $S_{DF}^*(f),S_{CF}^*(f),C_{CF}^*(f)$ and optimal Lagrange multipliers $\left(\lambda_C^*,\lambda_S^*\right)$ must satisfy the following KKT conditions
\begin{flalign}
\label{kkt_conditions}
&\nabla{L(f,S_{DF}^*(f),S_{CF}^*(f),C_{CF}^*(f),\lambda_C^*,\lambda_S^*)}=(0,0,0)\\
\nonumber&\lambda_S^*\left[\int_0^W[T_{DF}(f) \cdot S_{DF}^*(f) + T_{CF}(f) \cdot S_{CF}^*(f)] \cdot df - P\right]=0\\
\nonumber&\lambda_C^*\left[\int_0^W [T_{DF}(f) \cdot C_{DF}^*(f) + T_{CF}(f) \cdot C_{CF}^*(f)] \cdot df - C\right] =0\\
\nonumber&\int_0^W[T_{DF}(f) \cdot S_{DF}^*(f) + T_{CF}(f) \cdot S_{CF}^*(f)] \cdot df - P\leq0\\
\nonumber&\int_0^W [T_{DF}(f) \cdot C_{DF}^*(f) + T_{CF}(f) \cdot C_{CF}^*(f)] \cdot df - C\leq0\\
\nonumber&\lambda_C^*\geq0\\
\nonumber&\lambda_S^*\geq0
\end{flalign}
It can be readily seen that Slater's condition holds so we can solve the problem using the dual function. In Appendix \ref{DF_CF_appendix} we calculate $S_{DF}(f),S_{CF}(f),C_{CF}(f)$ solutions from the gradient condition of (\ref{kkt_conditions}) as a function of $\lambda_C,\lambda_S$. Those solutions require positive Lagrange multipliers values, therefore the power and rate constraints at (\ref{kkt_conditions}) would be equal to zero. Using the Lagrangian function from (\ref{lagrangian}) we next calculate the dual function
\begin{flalign}
\label{dual_function}
&g(\lambda_C,\lambda_S) =\\ &\nonumber max_{\substack{S_{DF}(f)\\S_{CF}(f)\\C_{CF}(f)}} L(f,S_{DF}(f),S_{CF}(f),C_{CF}(f),\lambda_C,\lambda_S)
\end{flalign}
In Appendix \ref{DF_CF_appendix} we find the solution of $S_{DF}(f), S_{CF}(f), C_{CF}(f)$ for the optimization problem in (\ref{dual_function}). We next substitute those solutions into (\ref{dual_function}) so this function of only $(\lambda_C,\lambda_S)$ would next be maximized in order to obtain the optimal solution for our problem
\begin{flalign}
\label{dual_function_maximize}
max_{\substack{\lambda_C,\lambda_S}}g(\lambda_C,\lambda_S)
\end{flalign}
Using Eq. (\ref{CF_multipliers_region}) and Eq. (\ref{DF_multipliers_region}) from Appendix \ref{DF_CF_appendix} we can derive bounds of the Lagrange multipliers that give an outer bound for the required region - the region where both CF and DF are feasible in which we would get a time sharing solution. However, this outer bound is not necessarily a region with only time sharing solutions to the problem, as it could also contain regions where either CF or DF is feasible. This is shown in Fig. \ref{lagrange_region_h_1}. For the optimal pair of $\lambda_S$ and $\lambda_C$ the equations in Appendix \ref{DF_CF_appendix} provide $S_{CF}(f), C_{CF}(f), S_{DF}(f), C_{DF}(f)$, but not $T_{CF}(f)$ and $T_{DF}(f)$. Using those powers and rates values we now optimize $T_{CF}(f)$ and $T_{DF}(f)$ by using linear programming (LP) methods. Their optimal solution must satisfy the total power and rate constraints. The LP problem for N frequency bands such that $N \cdot \Delta_f = W$ is described in Eq. (\ref{LP_problem}).
\begin{flalign}
\label{LP_problem}
&max\ \sum_{i=1}^{N} [T_{DF}(i) \cdot R_{DF}(i) + T_{CF}(i) \cdot R_{CF}(i)] \cdot 2\Delta_f \\
\nonumber&s.t.\\
\nonumber&0 \leq T_{DF}(i) \leq 1\\
\nonumber&0 \leq T_{CF}(i) \leq 1\\
\nonumber&0 \leq T_{DF}(i)+T_{CF}(i) \leq 1\\
\nonumber&0 \leq \sum_{i=1}^{N} [T_{DF}(i) \cdot S_{DF}(i) + T_{CF}(i) \cdot S_{CF}(i)] \cdot \Delta_f \leq P\\
\nonumber&0 \leq \sum_{i=1}^{N} [T_{DF}(i) \cdot C_{DF}(i) + T_{CF}(i) \cdot C_{CF}(i)] \cdot \Delta_f \leq C
\end{flalign}
\\
We will now summarize the optimization procedure
\begin{enumerate}
    \item Set the total power and rate constraints P and C, and the frequency dependent filter values. 
    \item For each point on the grid of all possible $\lambda_C$ and $\lambda_S$, first calculate optimal values of $S_{CF}(f), C_{CF}(f), S_{DF}(f), C_{DF}(f)$ for each frequency using the solution written in Appendix \ref{DF_CF_appendix}.
    \item Using those values, solve the LP problem in order to find optimal values of $T_{CF}(f)$ and $T_{DF}(f)$ so the time allocation for CF and DF will satisfy the total power and rate constraints.
    \item Calculate the system rate for the optimal values found above.
    \item Choose the grid point that maximizes the system rate.
\end{enumerate}
\subsection{Results}
\begin{figure}[htbp]
\centering
\includegraphics[width=0.5\textwidth]{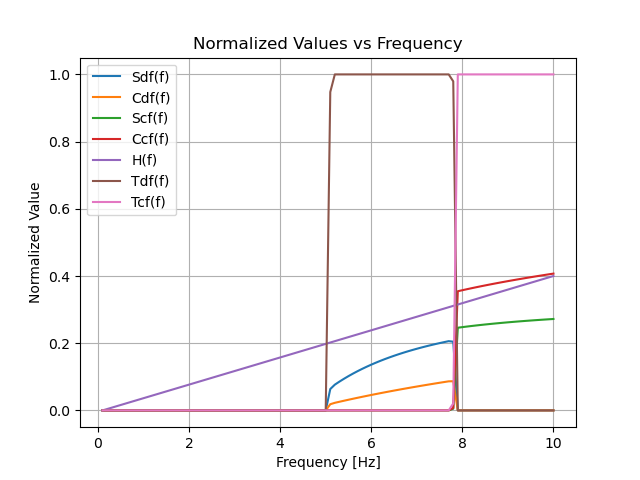}
\caption{Monotonically increasing filter frequency allocation}
\label{linear_filter_allocation}
\end{figure}

\begin{figure}[htbp]
\centering
\includegraphics[width=0.5\textwidth]{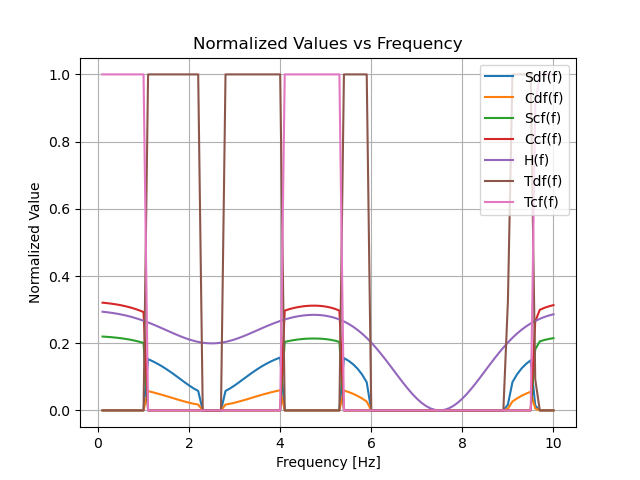}
\caption{Frequency-selective filter allocation}
\label{frequency_selective_filter_allocation}
\end{figure}

In this section we will show some results of optimal allocation. The frequency selective case optimization was done using Python Scipy.
First we examine a channel response monotonically increasing with frequency. With bandwidth of W=10[Hz] and with power and rate constraints P=100 and C=9 the allocation result is shown in Fig. \ref{linear_filter_allocation}.
As can be seen, the domain is devided into 3 regions. The first region with low filter values has no allocation. Second region has only DF allocation. Third region has only CF allocation. Between the regions there are two points of time sharing, first point does partly DF and second point does time sharing between CF and DF. This behavior corresponds to region 1 of Proposition 1.
In Fig. \ref{frequency_selective_filter_allocation} we show the allocation for bandwidth of W=10[Hz] with the filter used in \cite{our_prev} with P=100 and C=9.
As can be seen the optimal solution shows that for each frequency we choose either CF, DF or none and the partition between the regions would be at a specific filter value. The total system rate achieved with optimal time sharing between CF and DF is 7.5, compared to a lower value of 6.8 achieved with only CF in \cite{our_prev}. This is closer to the oblivious collaborative encoding upper bound rate of 8.16 shown in \cite{our_prev}. The optimal solution allocates power and fronthaul rate in frequency bands where CF only solution in \cite{our_prev} would not allocate, this is where the filter value is lower. Compared to the CF only solution, in the time-sharing solution the normalized rate allocation of CF is higher and the normalized power allocation of CF is lower so they do not get similar values as shown in \cite{our_prev}.
\section{Optimal solution properties}
\label{optimal_solution_section}
Next we analyze the behavior of the optimal solution for the frequency selective case.
\\
\textbf{Lemma 1:}\ The solution for each Lagrange multipliers point divides the channel frequency bands into two types according to the filter value at each band and a filter value threshold $H_{TH}$.
\begin{enumerate}
    \item Where $S_{DF}>S_{CF}$ for $H(f)<H_{TH}$.
    \item Where $S_{CF}>S_{DF}$ for $H(f)>H_{TH}$.
\end{enumerate}
\textbf{Proof:}\ The proof is shown in Appendix \ref{proof_lemma1}.\\
\begin{figure}[htbp]
\centering
\includegraphics[width=0.5\textwidth]{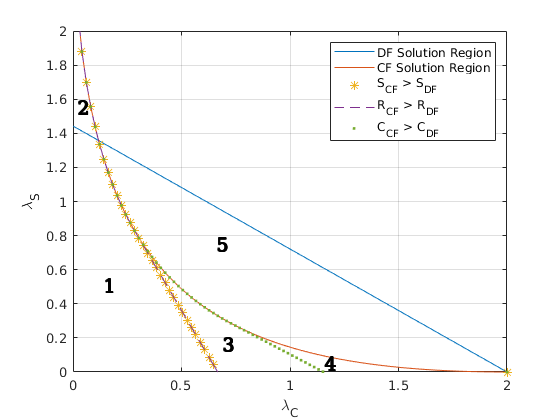}
\caption{Regions of CF and DF solutions on the Lagrange multipliers grid. Below the lines is the region where $S_{CF}>S_{DF}$.}
\label{lagrange_region_h_1}
\end{figure}
In Fig. \ref{lagrange_region_h_1} we show the regions border lines on the Lagrange multipliers grid. Below the CF and DF solution region lines are the regions where the solutions are feasible, where we define a non feasible solution if it is either negative or imaginary. Below the power, relay rate and channel rate lines is where CF has higher value. We now describe them according to the numbers shown on the figure.
Solution regions denoted by bold numbers in Fig. \ref{lagrange_region_h_1}:
\begin{enumerate}
    \item This region is where both CF and DF are allocated and CF has higher power, channel rate and relay rate.
    \item In this region only CF can be allocated,
    \item Here CF has higher relay rate.
    \item DF has higher power, channel rate and relay rate.
    \item Only DF is allocated.
\end{enumerate}
We can see in Fig. \ref{lagrange_region_h_1} that CF consumes less relay rate on the region lines of power (yellow) and system rate (green), therefore the optimal solution would use CF there. In region 3 one can easily show that DF is preferable by assigning in its channel rate equation power of $S_{CF}$ which is lower than $S_{DF}$ there. This approach does not change the channel rate region, thus allowing us to compare the solutions by only the required relay rate. DF relay rate is smaller in this region, therefore the optimal solution would prefer it there. The lines equations are calculated in Appendix \ref{solution_lines_appendix}.
Next we refer to the case of two frequencies with optimal allocation, and let suppose that in each one of them there is CF and DF part. To generalize this we will divide each frequency into part A and part B, each could be either DF or CF. This is shown in Fig. \ref{two_frequencies_time_sharing_replace}.
Theorem 2 shows the optimal solution behavior in this case.\\
\textbf{Theorem 2:} \ We define A and B to denote CF and DF respectively or in reverse order, that is, A may be CF and B denotes DF or A may be DF and then B denotes CF. We also define $\epsilon>0$ as $H(f_2)=(1+\epsilon)H(f_1)$ and $K>(1+\epsilon)^2$ as $S_{A}(f_1)=K\cdot S_{B}(f_2)$.
\\Then for two frequency bands with different filter values such that $H_2=H(f_2)>H(f_1)=H_1$ and $S_{A}(f_1)>S_{B}(f_2)$ and if there is some time in $f_2$ allocated to B, then A is not allocated in $f_1$ in the optimal solution.
\\
\textbf{Proof:} \ The proof is shown in Appendix \ref{proof_theorem1}. 
\\Using the above results we now state Proposition 1.\\
\textbf{Proposition 1:}\ Let the channel filter $H(f)$ be continuos in $f$. As was stated in Lemma 1, the optimal ($\lambda_C,\lambda_S$) point divide the filter values into 2 regions. The optimal allocation in those regions will be
\begin{enumerate}
    \item Region of higher $H(f)$ values, where $S_{CF}>S_{DF}$ and CF will be allocated at the higher channel gains and DF at the lower ones. Also a band of lowest channel gains may be left unused.
    \item Region of lower $H(f)$ where $S_{CF}<S_{DF}$ and only DF will be allocated. A band of lowest channel gains may be left unused.
\end{enumerate}
\textbf{Proof:} \ For any infinitesimally close channel gains the $\epsilon$ condition on $K$ is fulfilled. Then, by Theorem 2, in the region of higher $H(f)$ any CF allocation will migrate to all higher gain frequencies which have an DF allocation. In the lower $H(f)$ region DF is prefered as shown above with assigning it with CF power.

\section{Conclusions and outlook}
\label{section_conclusions}
In this work the band limited symmetric primitive diamond relay channel is considered, where a single user is connected to two non-cooperating relay nodes via symmetric bandlimited and filtered Gaussian channels, while the relay nodes are connected to the final-end receiver via ideal fronthaul links of given capacity. We consider and optimize achievable schemes that account for decode and forward (DF), and distributed compress and forward (CF), and compare the achievable rates to the cut-set bound and the upper bound from \cite{newupperbound}. We attempted to combine CF and DF by superposition in the same time and frequency resource similar to the classical scheme used for broadcast channels and proved that this approach yield no improvement over time-sharing. From the above discussion we can conclude that for frequency dependent filter with allocation for both CF and DF, higher filter values would prefer CF allocation and lower filter values would prefer DF allocation. We show that using CF and DF time sharing we can increase the total system rate relative to using only one of them. Frequency bands with filter value between CF and DF allocation would have time sharing and other frequencies would have either CF, DF or none. We can also view the flat frequency response model with time-sharing optimization to be equivalent to the uniform filter case on the frequency-selective model, where each frequency does either CF or DF. The time-sharing optimization shown for the flat frequency response is equivalent to frequency-sharing optimization, that is, part of the uniform bandwidth uses CF and the other part uses DF. The optimal values of those parts would be equal to the values of the optimal time-sharing for the flat frequency response model, taking into account the appropriate powers and rates constraints. Therefore, the diamond network optimization done in this paper is beneficial even for uniform filters, which is not the case where the classical water-pouring is used. Theoretical results addressing the primitive diamond channel may carry practical implications of future cell-free wireless technology \cite{compressionRAN}. For future work we suggest to examine whether binning can improve the achievable rates. In this case the relays decode part of the information by identifying the bin to which that information belongs, and the distributed compression is then aided by the bin identity, which is forwarded by both relays to the destination. This problem is open also in the discrete time primitive diamond channel (which is equivalent to a flat frequency response).

\appendices
\section{Optimal solution for CF and DF when $T_{CF}$ and $T_{DF}$ are given}
\label{DF_CF_appendix}
Here we develop the Lagrange multipliers solutions for CF and DF allocation for the frequency dependent case. We denote $H(f)=H$ for simplicity. We first calculate the gradient equations using Eqs. (\ref{lagrangian}),(\ref{gradient_lagrangian}),(\ref{kkt_conditions}). In each derivative we get either $T_{CF}$ or $T_{DF}$ with all arguments, therefore we assume that $0<T_{CF}<1$ and $0<T_{DF}<1$ so we can cancel them out when deriving the equations where the gradient is zero.
The gradient equations for $S_{DF}, S_{CF}, C_{CF}$ are
\begin{flalign}
\label{lagrange_multipliers_gradient_equations}
&\frac{dL}{dS_{DF}} = 2 \cdot \frac{dR_{DF}}{dS_{DF}} - \lambda_S-\lambda_C \cdot \frac{|H|^2}{2\cdot ln(2) \cdot (1+S_{DF}|H|^2)}=0&&\\
\nonumber &\frac{dL}{dS_{CF}} = 2 \cdot \frac{dR_{CF}}{dS_{CF}} - \lambda_S=0\\
\nonumber &\frac{dL}{dC_{CF}} = 2 \cdot \frac{dR_{CF}}{dC_{CF}} - \lambda_C=0
\end{flalign}
We next find $S_{DF}, S_{CF}, C_{CF}$ solutions that depend on the Lagrange multipliers. From $S_{CF}$ and $C_{CF}$ equations we would derive the same solution that was calculated and investigated in \cite{our_prev}, where we get two solutions and choose the one for which the CF rate function is concave. This solution also gives the constraints written in Eq. (\ref{CF_multipliers_region}). 
\begin{flalign}
\label{CF_multipliers_region}
&0 \leq \lambda_C \leq 2\\
\nonumber&0 \leq \lambda_S \leq \frac{2|H(f)|^2}{ln(2)}
\end{flalign}
DF solution can be calculated from the first equation
\begin{flalign}
\label{DF_lagrange_solution}
&S_{DF} = \frac{2-\lambda_C}{2\lambda_S ln(2)}-\frac{1}{H^2}&&\\
\nonumber &C_{DF} = \frac{1}{2ln(2)}ln\left(\frac{H^2(2-\lambda_C)}{2\lambda_S ln(2)}\right)&&
\end{flalign}
From those solutions we can derive constraints for the Lagrange multipliers. From $C_{DF}$ solution we require that the expression inside the logarithm is positive. From the requirement that $S_{DF}>0$ we derive another constraint, so the constraints for DF are
\begin{flalign}
\label{DF_multipliers_region}
&0\leq \lambda_C < 2 \\
\nonumber &0<\lambda_S < \frac{H^2(2-\lambda_C)}{2ln(2)}
\end{flalign}

\section{Solution region lines equations}
\label{solution_lines_appendix}
Here we derive the region lines equations for general filter value, shown in Fig. \ref{lagrange_region_h_1} for $H=1$. The solution region of DF is where $S_{DF}$ has real and positive values. Using Eq. (\ref{DF_lagrange_solution}) we can easily find DF region line because only real solution is possible, therefore the border line is where $S_{DF} = 0$. This equation gives the line
\begin{flalign}
\label{DF_solution_region}
&\lambda_{S}^{DF Region} = \frac{H^2(2-\lambda_C)}{2ln(2)}
\end{flalign}
Now, using $S_{CF}$ solution as explained in Appendix \ref{DF_CF_appendix}, CF region is where $S_{CF}$ solution is real and positive, therefore its border line equation is where either it is negative or the expression inside the square root equals zero. Negative solution will be where the nominator 
\begin{flalign}
\nonumber
&H^2\lambda_C -2H^2+\lambda_S ln(2) - \sqrt{J}
\end{flalign}
is positive. However, it can be easily shown that 
\begin{flalign}
\nonumber
&(H^2\lambda_C -2H^2+\lambda_S ln(2))^2 > J
\end{flalign}
for every $\lambda_C>0,\lambda_S>0$. Thus, if 
\begin{flalign}
\nonumber
&H^2\lambda_C -2H^2+\lambda_S ln(2) > 0
\end{flalign}
the nominator will be positive. The region border where it occurs is 
\begin{flalign}
\label{negative_region_line}
&\lambda_S = \frac{H^2(2-\lambda_C)}{ln(2)}
\end{flalign}
Now for the square root expression, solving $J=0$ gives the lines
\begin{flalign}
\label{root_expression_line}
&\lambda_S = \frac{H^2}{ln(2)}\left(3\lambda_C+2 \pm 2\sqrt{2}\cdot \sqrt{\lambda_C(\lambda_C+2)}\right)
\end{flalign}
In order to choose the relevant line, we can eliminate the plus sign line by substituting $\lambda_C=1$ and get that for this line $\lambda_S$ value is greater than the upper bound of Eq. (\ref{CF_multipliers_region}).
Now using that
\begin{flalign}
\nonumber
&2-\lambda_C-\left(3\lambda_C+2 - 2\sqrt{2}\cdot \sqrt{\lambda_C(\lambda_C+2)}\right)=&&\\
\nonumber&2\sqrt{2}\cdot \sqrt{\lambda_C(\lambda_C+2)}-4\lambda_C=&&\\
\nonumber&4\lambda_C\left(\sqrt{\frac{\lambda_C(\lambda_C+2)}{2\lambda_C^2}}-1\right)&&
\end{flalign}
and that this expression is non negative for $0 \leq \lambda_C \leq 2$, we can conclude that the line of Eq. (\ref{negative_region_line}) is always above the minus sign line of Eq. (\ref{root_expression_line}), which implies that there is no region with real and negative solution. Therefore, the region line for CF solution is
\begin{flalign}
\label{CF_solution_region_line}
&\lambda_{S}^{CF \ Region \ line} = \frac{H^2}{ln(2)}\left(3\lambda_C+2 - 2\sqrt{2}\cdot \sqrt{\lambda_C(\lambda_C+2)}\right)
\end{flalign}
Next in order to find the line where $S_{CF}=S_{DF}$, we first find $S_{CF}$ value over CF solution region line. Taking into account that lower Lagrange multipliers value result in larger power and rate values, we can simplify our analytical examination using $S_{CF}$ value on the CF region line. This value is obtained by assign $\lambda_S = \lambda_{S}^{CF \ Region \ line}$ in $S_{CF}$ expression.
\begin{flalign}
\label{CF_power_solution_on_region_line}
&S_{CF}^{CF \ Region \ line} = \frac{2-\lambda_C}{4H^2(3\lambda_C+2-2\sqrt{2\lambda_C(\lambda_C+2)})}-\frac{1}{4H^2}
\end{flalign}
Using (\ref{CF_power_solution_on_region_line}) and $S_{DF}$ solution from (\ref{DF_lagrange_solution}), we can find the line where $S_{DF}=S_{CF}^{CF \ Region \ line}$.
\begin{flalign}
\label{CFregion_DF_equal_power}
&\lambda_{S}^{S_{DF}=S_{CF}^{CF \ Region \ line}} = && \\ \nonumber &\frac{H^2(2-\lambda_C)}{3ln(2)}\left[2-\frac{2-\lambda_C}{4\lambda_C+4-3\sqrt{2\lambda_C(2+\lambda_C)}}\right]&&
\end{flalign}
Above this line $S_{CF}^{CF \ Region \ line}>S_{DF}$ and for each $\lambda_C$ value the $S_{CF}$ values inside the CF region are larger than on the region line, therefore $S_{CF}>S_{DF}$ over this region. Below that line there are $S_{DF}$ values that are larger than $S_{CF}$. There we can analytically find the line where $S_{CF}=S_{DF}$, which is (\ref{CF_DF_equal_power}). Combining (\ref{CFregion_DF_equal_power}) and (\ref{CF_DF_equal_power}) we get the border line where $S_{CF}=S_{DF}$.
\begin{flalign}
\label{CF_DF_equal_power}
&\lambda_S^{S_{CF}=S_{DF}} = \frac{H^2(2-3\lambda_C)}{2ln(2)}
\end{flalign}
Using similar approach we can find $C_{CF}$ on CF region line
\begin{flalign}
\label{CF_rate_solution_on_region_line}
&C_{CF}^{CF \ Region \ line} = &&\\
&\nonumber\frac{ln\left(\frac{(4-\lambda_C^2)(3\lambda_C+2-2\sqrt{2\lambda_C(\lambda_C+2)})}{2\lambda_C(7\lambda_C^2+16\lambda_C+4-5\lambda_C\sqrt{2\lambda_C(\lambda_C+2)}-6\sqrt{2\lambda_C(\lambda_C+2)}}\right)}{2ln(2)}&&
\end{flalign}
Next we find the line where $C_{CF}^{CF \ Region \ line}=C_{DF}$
\begin{flalign}
\label{CFregion_DF_equal_rate}
&\lambda_{S}^{C_{DF}=C_{CF}^{CF \ Region \ line}} = && \\ \nonumber &\frac{H^2\lambda_C(7\lambda_C^2+16\lambda_C+4)-H^2\lambda_C(\frac{3}{2}\lambda_C+1)(5\lambda_C+6)}{ln(2)(\lambda_C+2)(3\lambda_C+2-2\sqrt{2\lambda_C(\lambda_C+2)}}&&\\
\nonumber &+\frac{H^2\lambda_C(5\lambda_C+6)}{2ln(2)(\lambda_C+2)} &&
\end{flalign}
Above this line $C_{CF}^{CF \ Region \ line}>C_{DF}$ and inside CF region $C_{CF}$ values are larger than on the region line, therefore $C_{CF}>C_{DF}$ over this region. Below that line there are $C_{DF}$ values that are larger than $C_{CF}$. There we can analytically find the line where $C_{CF}=C_{DF}$, which is (\ref{CF_DF_equal_rate}). Combining (\ref{CFregion_DF_equal_rate}) and (\ref{CF_DF_equal_rate}) we get the border line where $C_{CF}=C_{DF}$.
\begin{flalign}
\label{CF_DF_equal_rate}
&\lambda_S^{C_{CF}=C_{DF}} = \frac{H^2(3\lambda_C-2\sqrt{3}\lambda_C-2\sqrt{3}+4}{ln(2)}
\end{flalign}
Using (\ref{CF_power_solution_on_region_line}) and (\ref{CF_rate_solution_on_region_line}) we can obtain $R_{CF}^{CF Region line}$ and then find the line equation where it equals $R_{DF}$. This line as similar behavior as (\ref{CFregion_DF_equal_power}) and meets CF region line at the same point. Using that the system rate increases with power or relay rate increase, we can use similar explanation, used above for the power and rate regions, also for the system rate line. Calculating the line where $R_{CF}=R_{DF}$ gives (\ref{CF_DF_equal_power}). Therefore, the lines where $R_{CF}=R_{DF}$ and $S_{CF}=S_{DF}$ are coincident lines.

\section{Lemma 1 Proof}
\label{proof_lemma1}
From the Lagrange multipliers equations at Eq. (\ref{kkt_conditions}) we get
\begin{flalign}
\nonumber\lambda_S = 2\frac{dR_{CF}}{dS_{CF}}\\
\nonumber\lambda_C = 2\frac{dR_{CF}}{dC_{CF}}
\end{flalign}
And define
\begin{equation}
    \nonumber\lambda_W = 2\frac{dR_{DF}}{dS_{DF}}
\end{equation}
Now assuming we are at the optimal solution point, we examine two dT intervals, which can be on two different frequencies or on the same frequency. The first dT interval uses DF and the second uses CF. The solution optimality ensures that any modification would not improve system performance. Now we increase the DF power $S_{DF}$ by $\epsilon$. This will increase the rate of this interval by $\Delta_{DF}=\frac{1}{2}\epsilon\lambda_W$. Relay rate of $\frac{1}{2}\epsilon\lambda_W$ at each relay will be required to support this. Those resources increase in the DF part must be reduced from the CF part, reducing the CF rate by $\Delta_{CF}=\frac{1}{2}\epsilon\lambda_S + \frac{1}{4}\epsilon\lambda_C\lambda_W$. At the optimal point the gradient equals to zero, so small modifications would not change the system rate. Therefore, the rate change must satisfy $\Delta_{DF}=\Delta_{CF}$. So we get
\begin{equation}
\label{lagrange_multipliers_equation_proof}
    \lambda_W = \frac{\lambda_S}{1-\frac{1}{2}\lambda_C}
\end{equation}
We note that Eq. (\ref{lagrange_multipliers_equation_proof}) can be derived from Eq. (\ref{lagrange_multipliers_gradient_equations}). Both equations express the global coupling between the CF and DF allocations. Next using Eq. (\ref{df_rate_frequency}) we can calculate the power allocation for the DF part. The DF rate derivative is
\begin{equation}
    \nonumber\lambda_W = \frac{dR_{DF}}{dS_{DF}}=\frac{|H|^2}{2(1+S_{DF}|H|^2)\cdot ln(2)}
\end{equation}
Therefore the DF power is
\begin{equation}
    \nonumber S_{DF} = max\left(0,\frac{1}{2\lambda_W\cdot  ln(2)}-\frac{1}{|H|^2}\right)
\end{equation}
The CF part allocation is calculated as done in \cite{our_prev}. In order to complete the proof we now investigate the power allocation region where $S_{CF}>S_{DF}$ along with the rate allocation region where $C_{CF}>C_{DF}$ and system rate region where $R_{CF}>R_{DF}$.
\begin{figure}[htbp]
\centering
\includegraphics[width=0.5\textwidth]{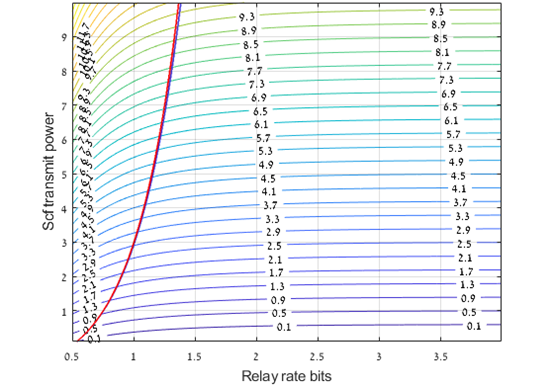}
\caption{Border lines of power and rate regions for different CF power and relay rate}
\label{power_rate_region}
\end{figure}
Fig. \ref{power_rate_region} shows the boundaries lines where $R_{CF}=R_{DF}$ (red) and $S_{CF}=S_{DF}$ (blue) on the CF power and relay rate grid. Above the lines is the region where DF has larger value. The contours show several $S_{DF}$ values. It is interesting to see that the lines coincide, as we will show next it creates strict boundary between the regions in each we prefer either CF or DF in the optimal allocation.
\begin{figure}[htbp]
\centering
\includegraphics[width=0.5\textwidth]{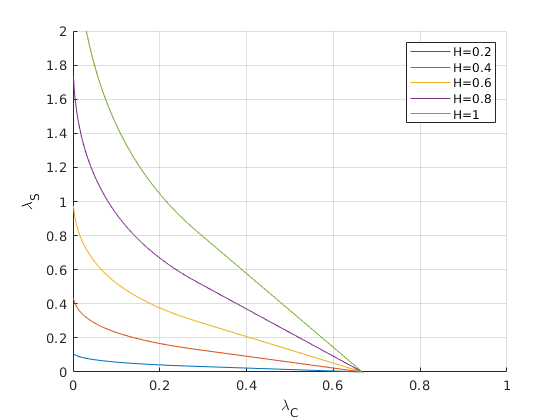}
\caption{Border lines of power region for different filter values on Lagrange multipliers grid}
\label{lagrange_region_filter_value}
\end{figure}
In Fig. \ref{lagrange_region_filter_value} we can see the power domain border as a function of the filter values. Below the lines is the region where $S_{CF}>S_{DF}$. Optimal solution is a certain point on this grid. It can be seen that the region gets smaller as the filter value reduces. This behavior can explain a prefered allocation for DF at lower filter values and  CF allocation for larger filter values. Each Lagrange multipliers point lies on a single border line which is defined by a specific filter value. By defining this filter value as $H(f)=H_{TH}$ we complete the proof.

\section{Theorem 2 Proof}
\label{proof_theorem1}
\begin{figure}[htbp]
\centering
\includegraphics[width=0.5\textwidth]{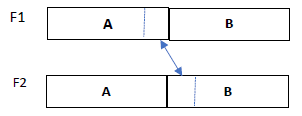}
\caption{Two frequencies time portion replace}
\label{two_frequencies_time_sharing_replace}
\end{figure}

Suppose we move a time portion of A from $f_1$ to $f_2$ and move corresponding duration time portion of B from $f_2$ to $f_1$ as shown in Fig. \ref{two_frequencies_time_sharing_replace}. 
We adjust the transmitted powers of A and B shifted portions such that the channel output power is preserved so the system throughput and the relays bit rate are conserved.
Therefore we get the following equations where the tag mentions the power at the other band.
\begin{flalign}
\nonumber &H_2^2S_{B2}=H_1^2S_{B2'} &&
\\\nonumber&\Rightarrow\Delta S_{B2'}=\left[\left(\frac{H_2}{H_1}\right)^2-1\right]S_{B2}=[2\epsilon+\epsilon^2]S_{B2}
\\\nonumber&H_1^2S_{A1}=H_2^2S_{A1'}
\\ \nonumber&\Rightarrow-\Delta S_{A1'}=\left[1-\left(\frac{H_2}{H_1}\right)^2\right]S_{A1}=\left[1-\frac{1}{(1+\epsilon)^2}\right]S_{A1}
\\\nonumber&=\frac{2\epsilon+\epsilon^2}{(1+\epsilon)^2}S_{A1}>\frac{2\epsilon+\epsilon^2}{(1+\epsilon)^2}K \cdot S_{B2}>\frac{2\epsilon+\epsilon^2}{(1+\epsilon)^2}(1+\epsilon)^2 \cdot S_{B2}
\\\nonumber&=[2\epsilon+\epsilon^2]S_{B2}=\Delta S_{B2'}
\end{flalign}

The inequality means that we have positive power remaining for allocation from this process. Using this excess power in the CF part we can increase the system throughput. However, this is a contradiction to the optimal point assumption.

\section*{Acknowledgment}
This work has been supported by the European Union's Horizon 2020 Research
And Innovation Program, grant agreement no. 694630, and the WIN consortium via the Israel minister of economy and science.
\bibliographystyle{IEEEtran}
\bibliography{IEEEabrv,citations}

\end{document}